\documentclass[aps, pra, reprint, superscriptaddress]{revtex4-1}
\usepackage{amssymb}
\usepackage{amsmath}
\usepackage{graphicx}
\usepackage{dcolumn}
\usepackage{bm}

\begin{document}

\title{Spectroscopic study on hot-electron transport in a quantum Hall edge
channel}
\author{Tomoaki Ota}
\thanks{TO and SA contributed equally to this work.}
\author{Shunya Akiyama}
\thanks{TO and SA contributed equally to this work.}
\affiliation{Department of Physics, Tokyo Institute of Technology, 2-12-1 Ookayama,
Meguro, Tokyo, 152-8551, Japan.}
\author{Masayuki Hashisaka}
\affiliation{Department of Physics, Tokyo Institute of Technology, 2-12-1 Ookayama,
Meguro, Tokyo, 152-8551, Japan.}
\affiliation{NTT Basic Research Laboratories, NTT Corporation, 3-1 Morinosato-Wakamiya,
Atsugi, 243-0198, Japan.}
\author{Koji Muraki}
\affiliation{NTT Basic Research Laboratories, NTT Corporation, 3-1 Morinosato-Wakamiya,
Atsugi, 243-0198, Japan.}
\author{Toshimasa Fujisawa}
\email{fujisawa@phys.titech.ac.jp}
\affiliation{Department of Physics, Tokyo Institute of Technology, 2-12-1 Ookayama,
Meguro, Tokyo, 152-8551, Japan.}

\begin{abstract}
Hot electron transport in a quantum Hall edge channel of an AlGaAs/GaAs
heterostructure is studied by investigating the energy distribution function
in the channel. Ballistic hot-electron transport, its optical-phonon
replicas, weak electron-electron scattering, and electron-hole excitation in
the Fermi sea are clearly identified in the energy spectra. The
optical-phonon scattering is analyzed to evaluate the edge potential
profile. We find that the electron-electron scattering is significantly
suppressed with increasing the hot-electron's energy well above the Fermi
energy. This can be understood with suppressed Coulomb potential with longer
distance for higher energy. The results suggest that the relaxation can be
suppressed further by softening the edge potential. This is essential for
studying non-interacting chiral transport over a long distance.
\end{abstract}

\date{\today }
\maketitle

\section{Introduction}

Hot electrons with the energy greater than the Fermi energy are subject to
relaxation processes such as electron-electron and electron-phonon
scattering \cite%
{BookRidley,EE2D-Giuliani,EE2D-negativeR,EE1D-Karzig,EP2D-Sivan,EPlowD-Bockelmann}%
. Therefore, ballistic and coherent electron transport is usually expected
only at low temperatures and low-energy excitation \cite%
{BookNazarov,hotEB=0-Rossler}. This also applies to chiral edge channels in
the integer quantum Hall regime \cite{QHMZ-YJi,QHHOM-Freulon,QHMZee-Tewari}.
While the conductance is quantized due to the absence of backscattering,
forward scattering is so significant that electronic excitation easily
relaxes to collective excitations in the plasmon modes \cite%
{leSueurPRL2010,HashisakaNatPhys}. This relaxation length is only a few $\mu 
$m when a small excitation energy of about 30 $\mu $eV is used for a GaAs
heterostructure, and decreases with increasing energy in agreement with the
spin-charge separation in the Tomonaga-Luttinger model \cite%
{leSueurPRL2010,TwoStage-Itoh}. However, recent experiments using a depleted
edge \cite{HotEDyQD-Kataoka} or a high-magnetic field \cite%
{HotEDyQD-Fletcher,HotEDyQD-Ubbelohde} have demonstrated ballistic transport
over 1 mm for hot electrons with surprisingly large energy of about 100 meV
above the Fermi energy \cite{HotEDyQD-Johnson}. In this high-energy region
greater than the optical phonon energy, the optical-phonon scattering
process has been studied extensively. The relaxation in the intermediate
energy region is yet to be investigated for how the electron-electron
scattering has been suppressed. This is particularly important for realizing
coherent transport of hot electrons \cite{QHMZee-Tewari,HotE-coherence-STM},
as the coherency can be reduced with the electron-electron scattering if
exists. For the experiments on quantum Hall edges, most of the works were
devoted to study low energy excitation below 1 meV. Taubert et al. have
investigated electron-hole excitation in the Fermi sea, from which some
hydrodynamic effects as well as optical-phonon scattering are studied for
higher energy greater than 100 meV \cite%
{HotE-TaubertB=0,HotE-TaubertB>0,HotE-Taubert}. However, this
non-spectroscopic scheme is not convenient for the purpose. High-energy hot
electron can be excited with a dynamic quantum dot driven by high-frequency
voltage. This scheme is attractive for generating a single hot electron, but
not convenient for tuning the energy in the wide range of interest.
Systematic measurements with a spectroscopic scheme are highly desirable to
investigate the hot-electron transport.

In this work, hot-electron spectroscopy is employed, where hot electrons are
injected from a point contact (PC) to an edge channel and the electrons
after propagation are investigated by using an energy spectrometer made of a
similar PC. With fine tuning of gate voltages on injector and detector PCs,
we have investigated ballistic hot-electron transport, multiple emission of
optical phonons showing `phonon replicas', small energy reduction associated
with weak electron-electron scattering, and electron-hole plasma in the
Fermi sea. They are explained with electron-electron scattering and
electron-phonon scattering, which can be tuned with the soft edge potential.
This electric-field effect will be useful in designing one-dimensional
hot-electron circuits.

\section{Measurement scheme}

Consider a two-dimensional electron system (2DES) under a perpendicular
magnetic field $B$ in the $-z$ direction (to the back of the 2DES). Near the
edge of the 2DES, Landau levels increase with the soft edge potential along $%
x$ axis as illustrated in Fig. 1(a) for the lowest Landau levels (LLLs) with
spin up and down branches. Here, Landau-level filling factor $\nu $ in the
range of $1<\nu <2$ is considered as a simplest case to study. This energy
profile can be regarded as the energy - momentum ($k_{y}=eBx/\hbar $)
dispersion relation under the Landau gauge \cite{BookEzawa}. We focus on hot
spin-up electrons well above the chemical potential $\mu $.

\begin{figure}[tbp]
\begin{center}
\includegraphics[width = 3.15in]{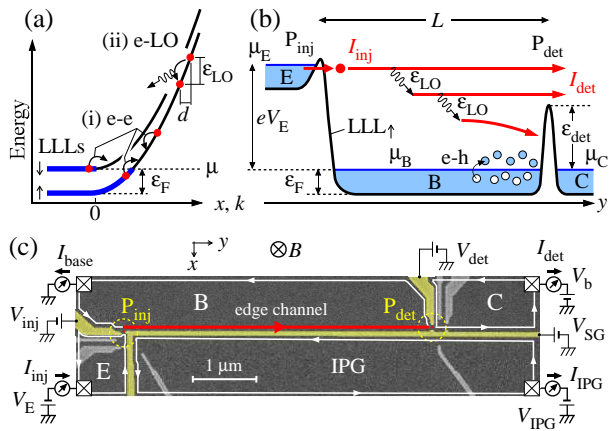}
\end{center}
\caption{(a) Schematic energy diagram of the lowest Landau levels (LLLs) for
spin-up an -down electrons. Electron-electron scattering (i) and LO phonon
scattering (ii) are illustrated. (b) Schematic energy diagram of hot
electron spectroscopy. A hot electron injected from emitter E experiences
relaxation associated with LO phonon emission and electron-electron
scattering. Resulting energy distribution function is measured with tunable
barrier $\protect\varepsilon _{\det }$. (c) The measurement setup with a
scanning electron micrograph of a $L=$ 5 $\protect\mu $m device. Other
devices have slightly different gate patterns but conceptually the same.
Bias voltage $-V_{\mathrm{E}}$ (with positive $V_{\mathrm{E}}$) is applied 
on the emitter contact to inject hot electrons to the edge
channel.}
\label{FIG1}
\end{figure}

The dominant relaxation processes in this system are also illustrated in
Fig. 1(a). Diagram (i) shows the electron-electron scattering between the
hot spin-up electron and cold electrons near the chemical potential \cite%
{LundePRB2010-EE,LundePRB-EE}. As the potential profile and thus the
dispersion is nonlinear, the two-particle scattering for exchanging equal
energy is basically forbidden as it cannot conserve the total momentum. The
scattering is practically allowed in the presence of random impurity
potential that breaks the translational invariance along $y$ axis. The
electron-electron scattering would be less probable for hotter electrons
with two reasons, as larger momentum mismatch and larger spatial separation
are involved. We investigate these effects in our experiment.

Diagram (ii) shows the optical-phonon scattering, where electron loses its
energy by emitting a longitudinal optical (LO) phonon with energy $%
\varepsilon _{\mathrm{LO}}=$ 36 meV in GaAs \cite{HotEDyQD-Kataoka}. This
phonon emission is suppressed by large spatial shift $d$ in the guiding
center of the electron motion, when $d$ is greater than the magnetic length $%
\ell _{\mathrm{B}}=\sqrt{\hbar /eB}$. We use this characteristics to
evaluate the effective electric field (or the potential profile) of the
channel. This provides better understanding of electron-electron scattering
as well as optical phonon scattering.

Figure 1(b) illustrates the measurement scheme for the hot electron
spectroscopy. The thick solid line labeled LLL$_{\uparrow }$ shows the
spin-up LLL (mostly in the bulk region) along the transport direction ($y$
axis). The injector PC labeled $\mathrm{P}_{\mathrm{inj}}$ and the detector
PC labelled \textrm{P}$_{\mathrm{\det }}$ separate three conductive regions;
the emitter (labeled E), the base (B), and the collector (C). These regions
are filled with electrons up to the respective chemical potentials, $\mu _{%
\mathrm{E}}$, $\mu _{\mathrm{B}}$, and $\mu _{\mathrm{C}}$, at the edges of
the conductive regions. With a large bias voltage $-V_{\mathrm{E}}$ on the
emitter, hot electrons with energy $eV_{\mathrm{E}}$ ($=\mu _{\mathrm{E}%
}-\mu _{\mathrm{B}}$) are injected from the emitter to the edge channel in
the base. Here, we assume that electrons are injected primarily into the
spin-up LLL in the base region, as tunneling to the spin-down LLL as well as
the second Landau levels (SLLs) is less probable with the thicker and higher
barriers.

In the base region, the hot electron loses its energy step by step by
emitting optical phonons and by generating electron-hole plasma in the Fermi
sea via electron-electron scattering. The resulting energy distribution
function is investigated with the detector \textrm{P}$_{\mathrm{\det }}$
located at distance $L$ from $\mathrm{P}_{\mathrm{inj}}$. Electrons with the
energy greater than barrier height $\varepsilon _{\det }$ are introduced to
the collector ($\mu _{\mathrm{C}}\simeq 0$), while other electrons with
lower energy are reflected and drained to the grounded based contact ($\mu _{%
\mathrm{B}}=0$). Therefore, the hot-electron spectroscopy can be performed
by measuring current $I_{\det }$ through \textrm{P}$_{\mathrm{\det }}$ at
various $\varepsilon _{\det }$.

Our measurement setup in the quantum Hall regime is shown in Fig. 1(c) with
a scanning electron micrograph of a test device. Surface metal gates
(colored yellow) were patterned on a modulation doped GaAs/AlGaAs
heterostructure (black). Magnetic field $B$ was applied perpendicular to the
heterostructure to form edge channels, and most of the measurements were
performed at bulk filling factor $\nu $ in the range of $1<\nu <2$. The main
edge channel (the red line) in the base is formed along the side gate SG.
The edge potential profile can be tuned with gate voltages $V_{\mathrm{SG}}$
on SG and $-V_{\mathrm{IPG}}$ on the other edge channel working as an
in-plane gate (IPG). Particularly, $V_{\mathrm{IPG}}=$ 0 - 0.2 V, with the
same sign of $V_{\mathrm{E}}$, is applied to eliminate the leakage of hot
electrons to the IPG. Tunneling barriers of the injector ($\mathrm{P}_{%
\mathrm{inj}}$) and the detector (\textrm{P}$_{\mathrm{\det }}$) were
adjusted by tuning voltages, $V_{\mathrm{inj}}$ and $V_{\mathrm{\det }}$,
respectively. Several devices with different $L$ = 0.7, 1.4, 5, 8, 10 and 15 
$\mu $m were formed with two-dimensional electron density $n_{\mathrm{2DES}%
}= $ 2.9$\times $10$^{11}$ cm$^{-2}$ (the zero-field Fermi energy of about
10 meV) and low-temperature mobility of $\mu _{\mathrm{2DES}}=$ 1.6$\times $%
10$^{6}$ cm$^{2}$/Vs (wafer W1) \cite{WashioPRB} or $n_{\mathrm{2DES}}=$ 2.6$%
\times $10$^{11}$ cm$^{-2}$ and $\mu _{\mathrm{2DES}}=$ 3$\times $10$^{6}$ cm%
$^{2}$/Vs (wafer W2). All measurements were performed at 1.5 - 2.1 K.

\section{Hot-electron spectra}

We measure the injection current $I_{\mathrm{inj}}$ and the detection
current $I_{\mathrm{\det }}$, which are defined as positive for forward
electron transport in the direction shown by the arrows in Fig. 1(c).
Ammeters with a relatively large input impedance of $Z_{\mathrm{m}}=$ 10 k$%
\Omega $ - 1 M$\Omega $ were used to prevent possible damage with unwanted
large current. The voltage drop in the ammeter is negligible for typical
current level of 0.1 - 1 nA, while some influences on measuring the
electron-hole plasma will be discussed later. The average number of injected
electrons travelling in the channel of length $L$, $I_{\mathrm{inj}}L/ev_{%
\mathrm{h}}$, is kept less than one, where $v_{\mathrm{h}}=E/B$ is the
hot-electron velocity for the electric field $E$ (discussed later) of the
edge potential, and thus the interaction between the injected electrons can
be neglected. The base current $I_{\mathrm{base}}$ at the base ohmic contact
and the leakage current $I_{\mathrm{IPG}}$ at IPG were always monitored to
ensure no leakage current ($I_{\mathrm{IPG}}=0$ and $I_{\mathrm{\det }}+I_{%
\mathrm{base}}=I_{\mathrm{inj}}$) within the noise level.

\begin{figure}[tbp]
\begin{center}
\includegraphics[width = 3.15in]{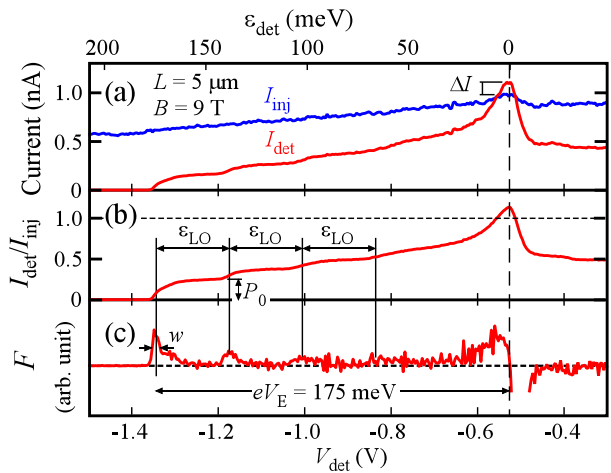}
\end{center}
\caption{(a) $I_{\mathrm{\det }}$ and $I_{\mathrm{inj}}$, (b) $I_{\mathrm{%
\det }}/I_{\mathrm{inj}}$, and (c) $F=d(I_{\mathrm{\det }}/I_{\mathrm{inj}%
})/dV_{\mathrm{\det }}$ as a function of $V_{\mathrm{det}}$ for a $L=$ 5 $%
\protect\mu $m device on wafer W1. The equispaced peaks in (c) represent the
ballistic transport (the leftmost peak) and its phonon replicas. The step
height $P_{0}$ in (b) measures the probability of ballistic transport for $%
L= $ 5 $\protect\mu $m. The peak with $I_{\mathrm{\det }}/I_{\mathrm{inj}}>1$
in (b) shows electron-hole excitation in the Fermi sea, and thus defines the
condition for $\protect\varepsilon _{\det }=0$. The width $w$ (= 5 meV in
energy) of the leftmost peak in (c) shows the energy resolution of this
measurement.}
\label{FIG2}
\end{figure}

Figure 2(a) shows a representative data of $I_{\mathrm{inj}}$ and $I_{%
\mathrm{\det }}$ at $V_{\mathrm{E}}=$ 175 mV as a function of $V_{\mathrm{%
\det }}$ for a $L$ = 5 $\mu $m device. As the injector \textrm{P}$_{\mathrm{%
inj}}$ with $I_{\mathrm{inj}}$ is slightly influenced by changing $V_{%
\mathrm{\det }}$, normalized current $I_{\mathrm{\det }}/I_{\mathrm{inj}}$
and its derivative $F=d(I_{\mathrm{\det }}/I_{\mathrm{inj}})/dV_{\mathrm{%
\det }}$ are evaluated as shown in Figs. 2(b) and 2(c), respectively. Here, $%
F$ is proportional to the energy distribution function of hot electrons in
the edge channel. The periodic stepwise increase of $I_{\mathrm{\det }}$
(peaks in $F$) manifests multiple LO phonon emissions. The width of the
peaks in $F$ is $w=$ 4 - 5 meV in energy, which is probably given by the
energy dependent tunneling probability in $\mathrm{P}_{\mathrm{inj}}$ and $%
\mathrm{P}_{\mathrm{\det }}$. This determines the energy resolution of the
spectroscopy. In the narrow region around $V_{\mathrm{\det }}\simeq -$0.55
V, the detector current exceeds the injection current ($I_{\mathrm{\det }%
}>I_{\mathrm{inj}}$), and the base current turns out to be negative ($I_{%
\mathrm{base}}<0$, not shown). This indicates electron-hole plasma in the
base, where the electrons with energy above $\varepsilon _{\det }$ and the
holes with energy below $\varepsilon _{\det }$ contribute excess detector
current. Therefore, the peak position in $I_{\mathrm{\det }}$ determines the
condition for $\varepsilon _{\mathrm{det}}=0$, where the top of the barrier
in $\mathrm{P}_{\mathrm{\det }}$ is aligned to $\mu _{\mathrm{B}}$ ($\simeq
\mu _{\mathrm{C}}$).

The energy scale of $\varepsilon _{\mathrm{det}}$ with respect to $V_{%
\mathrm{det}}$ is determined from the LO phonon replicas. For the data in
Fig. 2(a), linear dependence $\Delta \varepsilon _{\mathrm{det}}=\alpha
\Delta V_{\mathrm{det}}$ with the lever-arm factor $\alpha \simeq $ 0.213$e$
is confirmed from the equispaced LO phonon replicas. The spacing between the
leftmost peak in $F$ for the ballistic transport ($\varepsilon _{\mathrm{det}%
}=eV_{\mathrm{E}}$) and the zero energy peak ($\varepsilon _{\mathrm{det}}=0$%
) in $I_{\mathrm{\det }}$ is consistent with this $\alpha $. While some
devices showed nonlinearity in the $\varepsilon _{\mathrm{det}}$ - $V_{%
\mathrm{det}}$ relation, all spectroscopic analyses shown in this paper are
made with reasonable linearity.

A color plot of $F$ in Fig. 3, taken with $L=$ 1.4 $\mu $m device, captures
most of the features we discuss in this paper, where $V_{\mathrm{det}}$ is
converted to $\varepsilon _{\mathrm{det}}$ shown in the right axis. In the
high-energy region at $V_{\mathrm{E}}>$ 100 mV, the ballistic peak and its
phonon replicas are clearly resolved along the dashed lines ($\varepsilon _{%
\mathrm{det}}=eV_{\mathrm{E}}-n\varepsilon _{\mathrm{LO}}$ with $n=$ 0, 1,
and 2), which will be analyzed in Sec. IV. In the medium-energy region (30
mV $<V_{\mathrm{E}}<$ 60 meV), the highest-energy peak deviates from the
ballistic condition ($\varepsilon _{\mathrm{det}}=eV_{\mathrm{E}}$), which
will be explained with the weak electron-electron scattering in Sec. V. In
the low-energy region ($V_{\mathrm{E}}<$ 30 meV), no ballistic signal is
seen and the electron-hole excitation is clearly seen as a peak-and-dip
structure near $\varepsilon _{\mathrm{det}}=0$. This electron-hole plasma is
consistent with the weak electron-electron scattering as discussed in Sec.
V. In this way, the hot-electron spectroscopy is informative for analyzing
electron scattering.

\begin{figure}[tbp]
\begin{center}
\includegraphics[width = 3.15in]{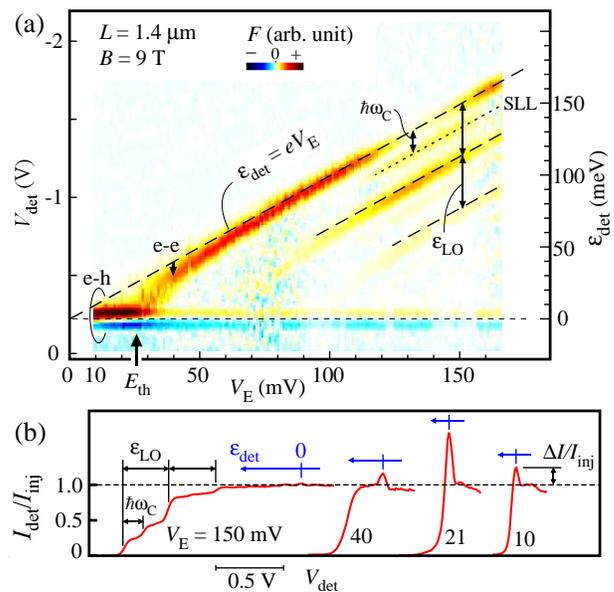}
\end{center}
\caption{Representative hot-electron spectrum, $F$ as a function of $\protect%
\varepsilon _{\det }$ on the right axis, for various $V_{E}$. The data is
taken at $B$ = 9 T ($\protect\nu =1.3$) with a $L=$ 1.4 $\protect\mu $m
device on wafer W2.}
\label{FIG3}
\end{figure}

\section{Optical-phonon scattering}

First, we analyze the optical-phonon scattering showing the phonon replicas
at $eV_{\mathrm{E}}>\varepsilon _{\mathrm{LO}}$ by ignoring the
electron-electron scattering. As shown in the inset to Fig. 4(a), the hot
electron in the LLL (the solid circle) can relax to a lower-energy state
(the open circle) via two possible processes; direct LO (dLO) phonon
emission within the LLL, and inter Landau level (iLL) tunneling to an
intermediate state (the open square) in the SLL followed by inter-LL LO
(iLO) phonon emission. Both can be dominant as studied in similar devices 
\cite{HotEDyQD-Johnson}. In our spectroscopic measurement, occupation in the
second Landau level (SLL) can be detected at a different condition, as the
barrier height for the SLL, $\varepsilon _{\mathrm{det}}+\hbar \omega _{%
\mathrm{C}}$, is higher than $\varepsilon _{\mathrm{det}}$ for the LLL. A
color-scale plot of $F$ in Fig. 4(a) shows such spectrum, where phonon
replicas of hot electrons in LLL (along the horizontal solid lines) and SLL
(along the dashed lines slanted by the cyclotron energy $\hbar \omega _{%
\mathrm{C}}$) are clearly seen. The peak spacing between the LLL and SLL
phonon replicas increases linearly with $B$ in agreement with the cyclotron
energy $\hbar \omega _{\mathrm{C}}$ [1.75 meV/T for GaAs]. This data shows
coexistence of the two relaxation processes in this sample. The iLL
tunneling may accompany acoustic phonon emission or absorption \cite%
{KomiyamaPRBspinflip}, but the corresponding phonon energy is too small to
be resolved in our measurement.

We find that this SLL signal appears only under some particular conditions
in some particular devices. We did not see systematic dependencies on $L$
and $V_{\mathrm{E}}$. While further studies are required, this implies that
the iLL tunneling is resonantly enhanced by an impurity or elsewhere. In
contrast, the LLL phonon replicas associated with the dLO process are
reproduced in various conditions. In the following, we analyze the LO phonon
scattering for the data without showing SLL signals.

\begin{figure}[tbp]
\begin{center}
\includegraphics[width = 3.15in]{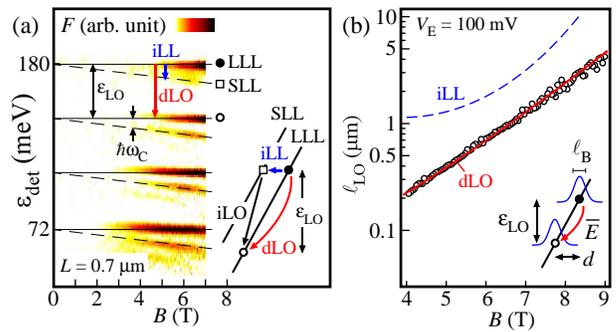}
\end{center}
\caption{(a) Phonon replicas of hot electrons in LLL and SLL seen in the
color-scale plot of $F$, taken at $V_{\mathrm{E}}=$ 180 mV and $V_{\mathrm{SG%
}}=$ -1.2 V with $L=$ 0.7 $\protect\mu $m device on wafer W2. The inset
shows the direct LO phonon emission (dLO) within the LLL, inter-Landau-level
tunneling (iLL) to the SLL, and inter-landau-level LO phonon emission (iLO).
(b) $B$ dependence of $\ell _{\mathrm{LO}}$, taken at $V_{\mathrm{E}}=$ 100
mV with a\ $L=$ 0.7 $\protect\mu $m device on wafer W2 [different from the
device in (a)]. The solid line is calculated for the dLO process, while the
dashed line is calculated for the iLL transition.}
\label{FIG4}
\end{figure}

For the dLO process, the LO phonon relaxation length $\ell _{\mathrm{LO}}$
is estimated from the probability $P_{0}=\exp \left( -L/\ell _{\mathrm{LO}%
}\right) $ of the ballistic transport for length $L$. Here $P_{0}$ is
directly obtained from the step height in the $I_{\mathrm{det}}/I_{\mathrm{%
inj}}$ trace [see $P_{0}$ in Fig. 2(b)] \cite{Pn}. As shown in Fig. 4(b), $%
\ell _{\mathrm{LO}}$ shows a clear exponential $B$ dependence. This can be
understood with the magnetic length $\ell _{\mathrm{B}}$ relative to the
spatial displacement $d$ between the initial and final states as shown in
the inset. When the edge potential is approximated by a linear $x$
dependence with average electric field $\bar{E}$ between the initial energy $%
\varepsilon $ and the final energy $\varepsilon -\varepsilon _{\mathrm{LO}}$%
, the displacement is given as $d=\varepsilon _{\mathrm{LO}}/e\bar{E}$, and
the LO phonon emission rate can be written as 
\begin{equation}
\Gamma _{\mathrm{LO}}=\Gamma _{\mathrm{LO,0}}\exp \left( -d^{2}/2\ell _{%
\mathrm{B}}^{2}\right)
\end{equation}%
where $\Gamma _{\mathrm{LO,0}}$ is the form factor that involves the
electron-phonon coupling constant in GaAs \cite%
{EPcalc-Telang,EPcalc-Emary2016,HotEDyQD-Johnson}. The corresponding
relaxation length is given by $\ell _{\mathrm{LO}}=v_{\mathrm{h}}/\Gamma _{%
\mathrm{LO}}$, where $v_{\mathrm{h}}=\bar{E}/B$ is the hot-electron
velocity. The data in Fig. 4(b) can be fitted well with this model at $\bar{E%
}$ = 1.13 MV/m and $\Gamma _{\mathrm{LO,0}}=$ 27 ps$^{-1}$ (the solid line
labelled dLO). If the relaxation were dominated by the iLL process, the
relaxation length should have had different $B$ dependence (the dashed line
labelled iLL \cite{Ebar4iLL}),\ as the tunneling distance $d_{\mathrm{iLL}%
}=\hbar \omega _{\mathrm{C}}/e\bar{E}$ for iLL depends on $B$. The observed
dependence in Fig. 4(b) suggests that the dLO process is dominant, and this
can be used to evaluate $\bar{E}$ in the edge potential.

\begin{figure}[tbp]
\begin{center}
\includegraphics[width = 3.15in]{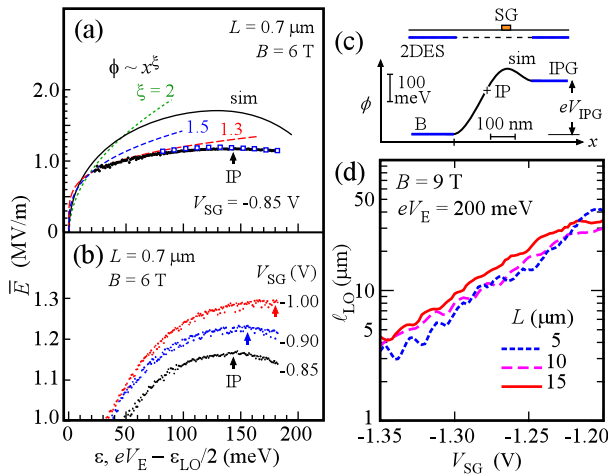}
\end{center}
\caption{(a and b) Energy $\protect\varepsilon $ dependence of $\bar{E}$.
The data with open squares were obtained from the $B$-dependence, while
other data with small dots were estimated from a single value of $\ell _{%
\mathrm{LO}}$. They were taken with a 0.7 $\protect\mu $m device on wafer
W2. The solid line in (a) is obtained from a self-consistent potential
calculation shown in (c). The dashed lines represents the electric field for
potential $\protect\phi \propto x^{\protect\xi }$ ($x>0$) with $\protect\xi %
= $ 2, 1.5, and 1.3. The lower-energy data follows $\protect\xi =$ 1.3. (c)
Self-consistent potential profile $\protect\phi $ for a realistic device
geometry, shown in the upper inset, with $V_{\mathrm{SG}}=$ -0.85 V and $V_{%
\mathrm{IPG}}=$ -0.2 V. (d) $V_{\mathrm{SG}}$ dependence of $\ell _{\mathrm{%
LO}}$, taken with three devices ($L=$ 5, 10, and 15 $\protect\mu $m) on
wafer W1.}
\label{FIG5}
\end{figure}

The energy ($eV_{\mathrm{E}}$) dependence of $\bar{E}$ is summarized in
Figs. 5(a) and (b), where $\bar{E}$ is plotted as a function of the average
energy $\bar{\varepsilon}=eV_{\mathrm{E}}-\varepsilon _{\mathrm{LO}}/2$ in
the dLO transition. The data with open squares in (a) were obtained from the 
$B$-dependence, while other data with small dots in (a) and (b) were
estimated from the measured $\ell _{\mathrm{LO}}$ and a fixed $\Gamma _{%
\mathrm{LO,0}}=$ 25 ps$^{-1}$. As clearly seen in the magnified plot of (b), 
$\bar{E}$ is maximized at $\bar{\varepsilon}\simeq $ 150 meV. This can be
understood with a realistic potential profile between the base region and
the IPG, as shown in Fig. 5(c). As the edge potential is defined by the
surface gate (SG), there must be an inflection point (IP) with the maximum
electric field. When $V_{\mathrm{SG}}$ is made less negative, electrostatics
suggests that $\bar{E}$ at each $\varepsilon $ as well as the IP position in 
$\varepsilon $ decrease, which is consistent with the data in Fig. 5(b).

Our data is compared to the calculated electrostatic potential. Here, we
have solved the Poisson equation with the boundary conditions around the
2DES and the gate. We used realistic device parameters; the 2DEG depth of
100 nm, the 2DEG thickness of 10 nm, the SG width of 80 nm, fixed surface
charge and ionized donor concentrations that produce the surface potential
and $n_{\mathrm{2DEG}}$, and applied voltages ($V_{\mathrm{SG}}$ = -0.85 V
and $V_{\mathrm{IPG}}$ = 0.2 V). The obtained potential profile $\phi \left(
x\right) $ is shown in Fig. 5(c), and its electric field is plotted with the
solid line labeled `sim' in Fig. 5(a). The simulation shows an IP at $%
\varepsilon \simeq $ 130 meV comparable to the measured one. The calculated
electric field is somewhat greater than the experimental values, possibly
due to imperfection of the model. 

Figure 5(a) shows quite weak
energy dependency of $\bar{E}$. Since $\bar{E}$ should be close to zero at
zero energy (0.08 MV/m in Ref. \cite{McClurePRL2009}), there must be a
drastic change of $\bar{E}$ in the low-energy region ($<$ 20
meV). While the edge potential $\phi $ is often approximated by a quadratic
form, this does not work well as shown by the dotted line (labelled $\xi =2$%
) for a quadratic potential with confinement energy of 5 meV. If we rely on
a fully 2D model neglecting the thickness of the heterostructure, the edge
potential has $\phi \propto x^{3/2}$ dependence in the lowest order near the
edge channel, as suggested from Eq. (7) and (8) in Ref. \cite{Chklovskii}.
Our experimental data implies that the potential in the low energy range (20 
$<$ $\varepsilon $ $<$ 60 meV) can be approximated
with a power dependence $\phi \propto x^{\xi }$ for $x>0$, as shown by the
dashed line with $\xi =1.3$. This energy dependency will be used in the
analysis of electron-electron scattering.\ 

For hot-electron applications, the LO phonon scattering can be suppressed by
decreasing $\bar{E}$, which can be done with less negative $V_{\mathrm{SG}}$
as seen in Fig. 5(b). Actually, $\ell _{\mathrm{LO}}$ reaches about 30 $\mu $%
m at $B$ = 9 T by tuning $V_{\mathrm{SG}}$, as shown in Fig. 5(d) taken with
several devices. Almost the same characteristics were reproduced with
different $L$, which ensures the validity of our measurements.

\section{Electron-electron scattering}

Next, we analyze the electron-electron interaction in the medium-energy
region. This part of the data in Fig. 2 is replotted in Fig. 6(a), where the
LO phonon scattering as well as the iLL process are not important. The
hot-electron signal is clearly visible at $eV_{\mathrm{E}}>E_{\mathrm{th}}$ (%
$\simeq $ 25 meV), while the electron-hole excitation near $\varepsilon
_{\det }=0$ is significant at $eV_{\mathrm{E}}<E_{\mathrm{th}}$. The latter
is characterized by the excess current $\Delta I$ obtained at $\varepsilon
_{\det }=0$ [see Fig. 2(a)]. Figure 6(b) shows the normalized excess current 
$\Delta I/I_{\mathrm{inj}}$ as a function of $V_{\mathrm{E}}$. It is
maximized at $eV_{\mathrm{E}}\simeq $ 25 meV, which coincides with the
vanishing point of the hot-electron signal. Therefore, we shall define $E_{%
\mathrm{th}}$ from the peak position in $\Delta I/I_{\mathrm{inj}}$. For $%
eV_{\mathrm{E}}<E_{\mathrm{th}}$, the hot electrons injected with energy $%
eV_{\mathrm{E}}$ are completely relaxed by exciting the Fermi sea, and the
lost energy $\delta =eV_{\mathrm{E}}$ should contribute finite $\Delta I/I_{%
\mathrm{inj}}$. For $eV_{\mathrm{E}}>E_{\mathrm{th}}$, the hot electrons are
partially relaxed by the energy loss $\delta $ ($<eV_{\mathrm{E}}$), which
should contribute $\Delta I/I_{\mathrm{inj}}$. Even at higher energy $eV_{%
\mathrm{E}}>$ 60 meV, the hot electron peak in Fig. 6(a) is slightly
deviated from the ballistic condition, and small but finite $\Delta I/I_{%
\mathrm{inj}}>0$ is seen in Fig. 6(b). They suggest the significance of
electron-electron scattering even for nominally ballistic hot-electron
transport.

For this problem, Lunde et al. have derived coupled Fokker-Plank equations
for distribution functions in the two channels \cite%
{LundePRB2010-EE,LundePRB-EE}. For simplicity, we focus only on the average
energy $\left\langle \varepsilon \right\rangle $ of hot electrons, provided
that the hot electrons are energetically separated from the Fermi sea. Then, 
$\left\langle \varepsilon \right\rangle $ follows a simple differential
equation 
\begin{equation}
\frac{d}{dy}\left\langle \varepsilon \right\rangle =-\gamma \left(
\left\langle \varepsilon \right\rangle \right) ,  \label{rate}
\end{equation}%
if each collision provides infinitesimal energy exchange. Here, $\gamma
\left( \varepsilon \right) $ is the energy relaxation rate per unit length
along $y$ direction. If $\gamma $ were independent of $\varepsilon $ as
assumed in Ref. \cite{LundePRB-EE}, the energy loss $\delta =\gamma L$ for a
fixed $L$ should have been independent of $\varepsilon $. Our result in Fig.
6(a) cannot be explained with a constant $\gamma $.

As we do not know the energy dependency of $\gamma \left( \varepsilon
\right) $ at this stage, we assume that $\gamma $ can be written as $\gamma
\left( \varepsilon \right) \simeq a\varepsilon ^{-\lambda }$ with parameters 
$\lambda $ and $a$. This form is convenient as this provides an analytical
solution of Eq. (\ref{rate}) and can be related to a physical model
described later. With initial energy $\left\langle \varepsilon \right\rangle
_{\mathrm{inj}}=eV_{\mathrm{E}}$, the final energy at $y=L$ follows 
\begin{equation}
\left\langle \varepsilon \right\rangle _{L}=\left[ \left( eV_{\mathrm{E}%
}\right) ^{\lambda +1}-E_{\mathrm{th}}^{\lambda +1}\right] ^{1/\left(
\lambda +1\right) },  \label{epsL}
\end{equation}%
where 
\begin{equation}
E_{\mathrm{th}}=\left[ \left( \lambda +1\right) aL\right] ^{1/\left( \lambda
+1\right) }  \label{EqEth}
\end{equation}%
is the threshold energy at which the hot electron just relaxes to the Fermi
level. Figure 6(c) shows some calculated traces $\left\langle \varepsilon
\right\rangle _{L}$ with $E_{\mathrm{th}}=$ 25 meV for several $\lambda =$
0.5, 1, 2, and 4. We find $\lambda =$ 1 $\sim $ 1.5 reproduces the
experimental data, as shown by the solid line with $\lambda =$ 1 overlaid in
Fig. 6(a).

The electron-hole excitation can be analyzed with the excess current $\Delta
I$ in Fig. 6(b). While the hot-electron spectroscopy works for energy
greater than $w\simeq $ 5 meV, electron-hole plasma in the Fermi sea is
distributed in a narrow energy range much smaller than $w$. Therefore, $%
\Delta I$ is based on thermoelectric current associated with the increased
temperature. For simplicity, we assume that the electron-hole plasma is
characterized by the Fermi distribution with an effective electron
temperature $T_{\mathrm{eff}}$, which is greater than the base temperature $%
T_{\mathrm{base}}$ in the collector \cite{TwoStage-Itoh,WashioPRB}. If the
lost energy $\delta $ is distributed to the two channels with a fraction $%
\beta $ for the spin-up channel ($\beta =\frac{1}{2}$ for equal energy
distribution), the corresponding heat power $W=\beta \delta I_{\mathrm{inj}%
}/e$ determines the effective temperature as
\begin{equation}
T_{\mathrm{eff}}^{2}-T_{\mathrm{base}}^{2}=\frac{6h}{\pi ^{2}k_{\mathrm{B}%
}^{2}}W
\end{equation}%
in the spin-up channel.
As $k_{\mathrm{B}}T_{\mathrm{eff}}$ is always smaller than $w$ in our
conditions, we can approximate that the tunneling probability of the
detector, $T\left( \varepsilon \right) \simeq T_{0}+\frac{\varepsilon }{2w}$%
, changes from $T_{0}$ ($=\frac{1}{2}$ at $\varepsilon _{\mathrm{det}}=0$)
linearly with small excess energy $\varepsilon $ ($\left\vert \varepsilon
\right\vert \ll w$) with respect to the chemical potential $\mu _{\mathrm{B}%
} $. With this model the thermoelectric current through $\mathrm{P}_{\det }$
follows%
\begin{equation}
I_{\mathrm{te}}\simeq \frac{\pi ^{2}}{12w}\frac{e}{h}k_{\mathrm{B}%
}^{2}\left( T_{\mathrm{eff}}^{2}-T_{\mathrm{base}}^{2}\right) ,  \label{EqDI}
\end{equation}%
for $k_{\mathrm{B}}T_{\mathrm{eff}}<w$. This yields the normalized thermal
current $I_{\mathrm{te}}/I_{\mathrm{inj}}\simeq \frac{1}{2w}\beta \delta $.

However, $\Delta I$ in the measurement should be smaller than $I_{\mathrm{te}%
}$ in the presence of series resistance in our setup. As shown in the
equivalent circuit between the base and the collector in the left inset to
Fig. 7(b), finite current $\Delta I$ induces voltage drop in the contact
resistance $R_{c}=\frac{h}{e^{2}}$ for the spin-up LLL and $Z_{\mathrm{m}}$
of the ammeters. A fraction of the thermoelectric current $I_{\mathrm{te}}$
flows back to the base through the tunneling resistance $R_{t}= \left( 
\frac{1}{T_{0}}-1\right) \frac{h}{e^{2}}$ of $\mathrm{P}_{\det }$ \cite%
{BookHeikkila}. The other fraction%
\begin{equation}
\eta =\frac{R_{t}}{R_{t}+R_{c}+2Z_{\mathrm{m}}}
\end{equation}%
of $I_{\mathrm{te}}$ is obtained in $\Delta I$ ($=\eta I_{\mathrm{te}}$)
with our setup. Therefore, we find a simple relation 
\begin{equation}
\Delta I/I_{\mathrm{inj}}=\frac{\eta \beta }{2w}\delta ,  \label{rel}
\end{equation}%
which relates $\Delta I/I_{\mathrm{inj}}$ in Fig. 6(b) and $\delta $ in Fig.
6(a). Note that this voltage drop is not important for hot electron
spectroscopy with a higher barrier ($\eta =1$ with $T_{0}=0$ for Fermi sea)
at $\varepsilon _{\mathrm{det}}\gg w$.

\begin{figure}[tbp]
\begin{center}
\includegraphics[width = 3.15in]{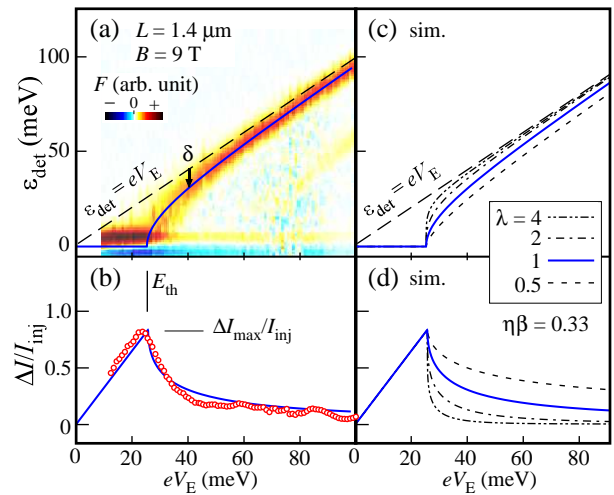}
\end{center}
\caption{(a) The color-scale plot of $F\left( \protect\varepsilon _{\det
},V_{E}\right) $. The peak is deviated from the ballistic condition (the
dashed line), and is explained well with the model calculation (the solid
line at $\protect\lambda =1$). (b) The obtained $\Delta I/I_{\mathrm{inj}}$
(solid circles) as a function of $eV_{E}$. The data is consistent with the
model calculation (the solid line at $\protect\lambda =1$) (c and d) The
model calculation of hot-electron energy $\left\langle \protect\varepsilon %
\right\rangle _{L}$ in (c) and $\Delta I/I_{\mathrm{inj}}$ in (d) for
different $\protect\lambda $.}
\end{figure}

If the average hot-electron energy follows Eq. (\ref{epsL}) and the lines in
Figs. 6(a) and 6(c), the corresponding $\Delta I/I_{\mathrm{inj}}$ should
follow the lines in Figs. 6(b) and 6(d).  Here, we chose $w=$ 5 meV and $%
\eta \beta =$ 0.33 to adjust the maximum $\Delta I/I_{\mathrm{inj}}$ to the
experimental one. The parameter $\eta \beta $ [0.2 - 0.5 in Fig. 7(b)] is
consistent with the equal heat distribution ($\beta =\frac{1}{2}$) and $\eta
\simeq $ 0.4 for $T_{0}=$ 0.5 and $Z_{\mathrm{m}}=$ 10 k$\Omega $. The
excellent agreement with the experimental data is found also in the high
energy tail at $eV_{\mathrm{E}}>$ 40 meV. Namely, the both data sets in
Figs. 6(a) and 6(b) are understood with the same energy loss $\delta $.

Figure 7 summarizes the $B$ dependence of $E_{\mathrm{th}}$ in (a) and the
normalized peak value $\Delta I_{\mathrm{\max }}/I_{\mathrm{inj}}$ ($\Delta
I/I_{\mathrm{inj}}$ at $\varepsilon =E_{\mathrm{th}}$) in (b) for several
devices. The threshold energy $E_{\mathrm{th}}$ does not change with $B$ in
Fig. 7(a). Weak $B$ dependence of $\Delta I_{\mathrm{\max }}/I_{\mathrm{inj}%
} $ is seen in Fig. 7(b). The $L$ dependence of $E_{\mathrm{th}}$ shown in
the inset to Fig. 7(a) is consistent with Eq. (\ref{EqEth}); $E_{\mathrm{th}%
}\propto L^{0.4}$ $\sim $ $L^{0.5}$ (the dashed line) with $\lambda =1-1.5$.

In a standard electron-electron scattering model, a hot spin-up electron can
relax by exchanging the energy with a cold spin-down electron or a cold
spin-up electron, as depicted in Fig. 1(a). The latter process may be
suppressed by the destructive interference with a similar process for
exchanged final states \cite{LundePRB2010-EE}, while such suppression should
be incomplete in the presence of energy dependent relaxation rate $\gamma
\left( \varepsilon \right) $. Nevertheless, spin-up and spin-down electrons
in their Fermi seas are easily thermalized by the proximate interaction \cite%
{leSueurPRL2010,TwoStage-Itoh}. This suggests equal heat distribution
between the two channels ($\beta =\frac{1}{2}$) in agreement with the
comparison in Fig. 6(b).

When the filling factor $\nu $ is increased above 2 ($B <$ 5.2
T), the heat can be distributed to electrons in the SLL. However, we did not
see such characteristics in Figs. 7(a) and 7(b). If the hot electron
scatters with electrons in the SLL [the dashed line in the right inset to
Fig. 7(a)], the excess scattering should increase $E_{\mathrm{th}}$ at $\nu
>2$. If the Fermi seas in the LLLs are interacting with electrons in the SLL
[the dashed line in the right inset to Fig. 7(b)], the heat redistribution
should decrease $\beta $ and thus $\Delta I_{\mathrm{\max }}/I_{\mathrm{inj}%
} $ at $\nu >2$. It seems both scattering processes with the SLL are
negligible for the short length ($<$ 1.4 $\mu $m), possibly due
to the large cyclotron energy that determines the channel distance between
SLL and LLL as compared to the small Zeeman energy that determines the
distance between spin-up and -down channels.

\begin{figure}[tbp]
\begin{center}
\includegraphics[width = 3.15in]{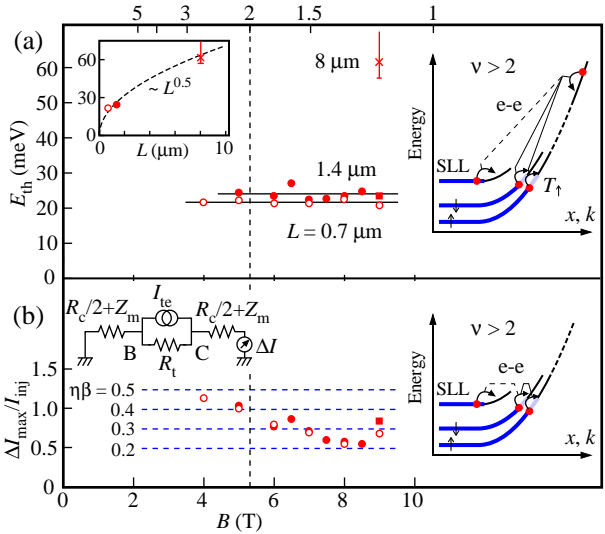}
\end{center}
\caption{$B$ dependence of $E_{\mathrm{th}}$ in (a) and $\Delta I_{\mathrm{%
\max }}/I_{\mathrm{inj}}$ in (b) taken with several devices on wafer W2. The
vertical dashed line shows the condition for $\protect\nu $ = 2. The inset
to (a) shows the interaction between a hot electron and electrons in the
LLLs and the SLL at $\protect\nu >$ 2. The right inset to (b)
shows the interaction between the LLLs and the SLL. The left inset to (b)
shows an equivalent circuit for relating $\Delta I$ with $I_{\mathrm{te}}$.}
\end{figure}

Now, we discuss the reason why the electron-electron scattering with $\gamma
\left( \varepsilon \right) $ is suppressed with increasing energy. The
electron-electron scattering should be sensitive to the potential profile $%
\phi \propto x^{\xi }$ ($\xi >0$) discussed with Fig. 5. A hot electron with
higher energy $\varepsilon $ is more spatially separated from the Fermi sea
(the distance $x\propto \varepsilon ^{1/\xi }$). This appears in the Coulomb
potential $U$ between the hot electron and an electron in the Fermi sea. If
we ignore the screening effect from the gate metal, the bare Coulomb
potential $U\propto x^{-1}\propto \varepsilon ^{-1/\xi }$ decreases with
increasing $\varepsilon $. Incidentally, the hot-electron velocity $v_{%
\mathrm{h}}$ is significantly greater than the Fermi velocity $v_{\mathrm{F}%
} $. With faster $v_{\mathrm{h}}$, the hot electron passes through the
channel with less scattering in a shorter time. Moreover, electron-electron
scattering should be suppressed with larger momentum mismatch proportional
to $\left\vert v_{\mathrm{F}}^{-1}-v_{\mathrm{h}}^{-1}\right\vert $. All of
these effects reduce the scattering of hot electrons with larger $%
\varepsilon $ and faster $v_{\mathrm{h}}$.

The scattering is allowed in the presence of random impurity potential,
which fluctuates the Coulomb potential $U$ around the mean $U_{0}$ with the
Fourier amplitude $A$ in the long-range limit over the correlation length $%
\ell _{p}$. In this case, $\gamma $ can be written as 
\begin{equation}
\gamma =\frac{\hbar U_{0}^{2}Av_{\mathrm{F}}}{4\sqrt{2}\pi ^{3/2}\ell
_{p}^{3}v_{\mathrm{h}}^{2}}  \label{ita}
\end{equation}%
in the limit of $v_{\mathrm{h}}\gg v_{\mathrm{F}}$, as derived in Eqs. (2)
and (9) of Ref. \cite{LundePRB-EE}. Since $A$, $\ell _{p}$, and $v_{\mathrm{F%
}}$ are irrelevant to the hot-electrons, $U_{0}$ ($\propto \varepsilon
^{-1/\xi }$) and $v_{\mathrm{h}}$ ($=E/B\propto \varepsilon ^{\left( \xi
-1\right) /\xi }$) suggest the energy dependency of $\gamma \left(
\varepsilon \right) \propto \varepsilon ^{-2}$. This exponent is close to
but somewhat larger than our experimental value of $\lambda =$ 1 $\sim $ 1.5
obtained for $\gamma \propto \varepsilon ^{-\lambda }$ in Fig. 6.

It should be noted that Eq. (\ref{ita}) does not explain the absence of $B$
dependence of $E_{\mathrm{th}}$ in Fig. 7(a). If $v_{\mathrm{h}}$ and $v_{%
\mathrm{F}}$ have $1/B$ dependence, we expect a measurable $B$ dependence in 
$E_{\mathrm{th}}\propto \left( v_{\mathrm{F}}/v_{\mathrm{h}}^{2}\right)
^{1/\left( \lambda +1\right) }$, which should exhibit $E_{\mathrm{th}%
}\propto B^{0.4}$ $\sim $ $B^{0.5}$ for $\lambda =$ 1 $\sim $ 1.5. The
discrepancy might be related to the formation of many-body states in LLLs.
At least, our previous work have shown that $v_{\mathrm{F}}$ is
significantly enhanced by the Coulomb interaction with the
Tomonaga-Luttinger model \cite{KumadaEMP,KamataNatNano2014}. Such many-body
states are not considered in the derivation of Eq. (\ref{ita}) \cite%
{LundePRB-EE}. A single-particle hot electron scattering with many-body
state may be worthy for studying non-linear hydrodynamic effect \cite%
{NonlinearTLL,1D2DTunSpectrum}.

For hot-electron applications, the electron-electron scattering can be
suppressed by decreasing $U_{0}$. This can be done with hotter electrons
with longer distance from the Fermi sea or with screening effect by covering
the surface with metal \cite{HotEDyQD-Kataoka}.

\section{Summary}

In summary, we have investigated hot electron transport in the soft edge
potential by means of hot electron spectroscopy. We find that the
electron-electron interaction is suppressed for hotter electrons. The
electron-phonon interaction is also suppressed by softening the edge
potential. The observed ballistic hot-electron transport is attractive for
utilizing hot electrons for studying electronic quantum optics.

\begin{acknowledgments}
We thank Yasuhiro Tokura for fruitful discussions. T.O. acknowledges the
financial support by Support Center for Advanced Telecommunications
Technology Research (SCAT). This work was supported by JSPS KAKENHI
(JP26247051, JP15H05854, JP17K18751), Nanotechnology Platform Program of
MEXT, Advanced Research Center for Quantum Physics and Nanoscience at Tokyo
Institute of Technology (TokyoTech), and Research Support Center for
Low-Temperature Science at TokyoTech. 
\end{acknowledgments}


\begin{thebibliography}{99}
\bibitem{Pn} The probability $P_{n}$ of transport after emitting just $n$
phonons can also be used to estimate the relaxation length $\ell _{\mathrm{LO%
}}^{\left( n\right) }$ of partially relaxed hot electrons at $eV_{\mathrm{E}%
}-n\varepsilon _{\mathrm{LO}}$. By solving the rate equations for the
cascaded LO phonon emissions, $\ell _{\mathrm{LO}}^{\left( n\right) }$ can
be estimated by using $\ell _{\mathrm{LO}}^{\left( n^{\prime }\right) }$ for 
$n^{^{\prime }}<n$. We find that $\ell _{\mathrm{LO}}^{\left( n\right) }$
does not contradict with $\ell _{\mathrm{LO}}$ obtained from $P_{0}$, if
they are compared for the same energy. However, the cascade analysis
involves large uncertainty associated with the error accumulation.
\bibitem{Ebar4iLL} The parameters are chosen to provide a comparable slope
to the experimental data in the $\ell _{\mathrm{LO}}-B$ plot, while the
significant bowing does not match the data.
\end{thebibliography}
\end{document}